\documentclass[prb,twocolumn,preprintnumbers,amsmath,amssymb,floats,citeautoscript,nobalancelastpage,showpacs]{revtex4-1}

\usepackage{graphicx}
\usepackage{bm}

\begin{document}

\title{Unconventional magnetism in the layered oxide LaSrRhO$_4$} 

\author{Noriyasu Furuta}
\author{Shinichiro Asai}
\author{Taichi Igarashi}
\author{Ryuji Okazaki}
\author{Yukio Yasui}
\altaffiliation[Present address:]{
Department of Physics, Meiji University, 
Kawasaki 214-8571, Japan
}
\author{Ichiro Terasaki}
\email[Email me at:]{
terra@cc.nagoya-u.ac.jp
}

\affiliation{Department of Physics, Nagoya University, Nagoya 464-8602, Japan}

\author{Masami Ikeda}
\author{Takahito Fujita}
\author{Masayuki Hagiwara}

\affiliation{KYOKUGEN
(Center for Quantum Science and Technology under Extreme Conditions),
Osaka University, Toyonaka, Osaka 560-8531, Japan}

\author{Kensuke Kobayashi}
\author{Reiji Kumai}
\author{Hironori Nakao}
\author{Youichi Murakami}

\affiliation{Condensed Matter Research Center and Photon Factory, 
Institute of Materials Structure Science, 
High Energy Accelerator Research Organization, Tsukuba 305-0801, Japan}

\begin{abstract}
 We have prepared polycrystalline samples of 
 LaSrRh$_{1-x}$Ga$_x$O$_4$ and LaSr$_{1-x}$Ca$_x$RhO$_4$,
 and have measured the x-ray diffraction,
resistivity, Seebeck coefficient, magnetization and electron spin resonance
 in order to evaluate their electronic states.
The energy gap evaluated from the resistivity and the Seebeck
 coefficient systematically changes with the Ga concentration,
and suggests that the system changes from a small polaron insulator 
to a band insulator.
 We find that all the samples show Curie-Weiss-like susceptibility with 
 a small Weiss temperature of the order of 1 K,
 which is seriously incompatible with the collective wisdom
 that a trivalent rhodium ion is nonmagnetic.
 We have determined the $g$ factor to be  $g$=2.3 from 
 the electron spin resonance, and the spin number to be $S$=1
 from the magnetization-field curves by fitting with a modified 
 Brillouin function.
 The fraction of the $S$=1 spins is 2--5\%, which depends 
 on the degree of disorder in the La/Sr/Ca-site, which implies
that disorder near the apical oxygen is related to the magnetism
of this system.
A possible origin for the magnetic Rh$^{3+}$ ions is discussed.
\end{abstract}


\maketitle
\section{Introduction}
The 3d transition-metal oxides have been extensively studied as a gold
mine for functional materials, which is exemplified by 
the ferroelectricity in titanium oxides, the magnetoresistivity and
multiferroelectricity in manganese oxides, 
the thermoelectricity in cobalt oxides, 
and the high-temperature  superconductivity in copper oxides.
In contrast,  the 4d transition metal oxides
have been less investigated as functional materials.
While magnetism is a fertile source for the functions in the $3d$
transition-metal oxides,
the $4d$ transition-metal oxides 
are often paramagnetic except for 
some insulating ruthenium oxides.\cite{subramanian1983,battle1984,nakatsuji1997}
This comes from different spin states between $3d$ and $4d$ elements.

The spin state is a fundamental concept
in transition-metal compounds/complexes \cite{sugano1970}.
In a transition-metal ion surrounded with 
octahedrally-coordinated oxygen anions, 
the five-fold degenerate $d$ orbitals
in vacuum are split into the triply degenerate 
$t_{2g}$ ($xy$, $yz$ and $zx$) orbitals and 
the doubly degenerate  $e_g$ ($x^2-y^2$ and $z^2$) orbitals, 
and the energy gap between the $t_{2g}$ and $e_g$ levels 
called the ligand-field gap often competes with the Hund coupling. 
When the ligand field gap is larger, the $d$ electrons 
first occupy the $t_{2g}$ states 
to minimize the total spin number.
On the other hand, when the Hund coupling
is strong, the total spin number is maximized.
The former state is called the low spin state, 
and the latter the high spin state.

Rhodium is located below cobalt in the periodic table, 
and thus is expected to have similar chemical properties. 
In fact, many cobalt oxides have their isomorphic rhodium oxides, 
and similar transport properties are reported.
\cite{okada2005,okada2005jpsj,klein2006,okamoto2006,shibasaki2006,maignan2009}
We have studied the Rh substitution effects on LaCoO$_3$,
and found that LaCo$_{0.8}$Rh$_{0.2}$O$_3$ exhibits a ferromagnetic
transition below 18 K.\cite{asai2011}
The substituted Rh ions tend 
to stabilize high-spin state Co$^{3+}$ ions in the samples,\cite{knizek2012,asai2012}
and such high-spin state Co$^{3+}$ ions interact with each other
at low temperatures to cause the ferromagnetic order.
This clearly indicates that the Rh$^{3+}$ ion is \textit{not} 
a simple nonmagnetic element.

In this paper, we focus on the layered rhodium oxide LaSrRhO$_4$.
This oxide crystallizes in the K$_2$NiF$_4$-type ($A_2B$O$_4$-type)
structure, where 50\% La and 50\% Sr make a solid solution in the $A$ site.
The corner-shared RhO$_6$ octahedra form the RhO$_2$
plane along the $ab$ plane, and alternately stack with the (La/Sr)$_2$O$_2$ layer.
 Shimura et al. \cite{shimura1994} measured the physical properties of
Sr$_{2-x}$La$_x$RhO$_4$ , and found that a small amount of paramagnetic
contribution, although the formal valence of Rh was $3+$.
This is highly unusual, because Rh$^{3+}$ is believed to be highly
stable in the low-spin state ($S=0$). 
Here we present measurements and analyses of the transport
and magnetic properties of LaSrRh$_{1-x}$Ga$_x$O$_4$
and LaSr$_{1-x}$Ca$_x$RhO$_4$,
and show that 2--5\% of the Rh$^{3+}$ ions act as $S=1$.
We discuss a possible mechanism to create the magnetic Rh$^{3+}$ ions
based on the experimental results.

\section{experimental}
Polycrystalline samples of LaSrRh$_{1-x}$Ga$_x$O$_4$ 
($x=$0, 0.1, 0.2, 0.3, 0.4, 0.5, 0.6, 0.7, 0.8 and 0.9)
and LaSr$_{1-x}$Ca$_x$RhO$_4$ ($x=$0.1, 0.2 and 0.3)
were prepared by a conventional solid state reaction method. 
High-purity (99.9\%) oxide powders of Rh$_2$O$_3$, 
Ga$_2$O$_3$, La$_2$O$_3$, SrCO$_3$  and CaCO$_3$ were used as raw materials.
Stoichiometric mixtures of these powders were ground, and were
calcined for 24 h at 1200$^{\circ}$C in air. 
The calcined powder was then ground, pressed into pellet, 
and sintered 48 h at 1300$^{\circ}$C in air.

X-ray diffraction was measured with a Rigaku Geigerflex 
(Cu K$\alpha$ radiation). 
Synchrotron x-ray diffraction was taken for a powder sample of 
$x=0.5$ with a wave length of 0.6887~\AA~ at BL-8A, KEK-PF, Japan.
Rietveld refinement was conducted using Rietan 2000 code.\cite{izumi2007}
Resistivity was measured in a four probe configuration in a constant 
voltage applied across a series circuit of a sample and a standard
resistance.
The Seebeck coefficient was measured in a two-probe configuration
in a steady state technique with a typical temperature gradient of 1
K/cm.
A contribution of the voltage leads was carefully subtracted. 
Magnetization was measured using 
a superconducting quantum interference device 
magnetometer (Quantum Design MPMS) from 5 to 300 K. 
External magnetic field $\mu_0 H$ was chosen to be from 1 to 5 T,
depending on the magnetization of the samples. 
The field dependence of the magnetization was measured for LaSrRhO$_4$
at 2, 5 and 10 K in sweeping $\mu_0 H$ from 0 to 7 T. 
Electron spin resonance (ESR) was measured in static 
magnetic fields from 0 to 14 T in the frequency range from
90 to 200 GHz, and non-resonant transmission signal was detected 
using a vector network analyzer in sweeping magnetic fields. 
For frequencies of 27.5 and 34 GHz, 
a cavity perturbation technique was employed.

\begin{figure}
 \includegraphics[width=6cm,clip]{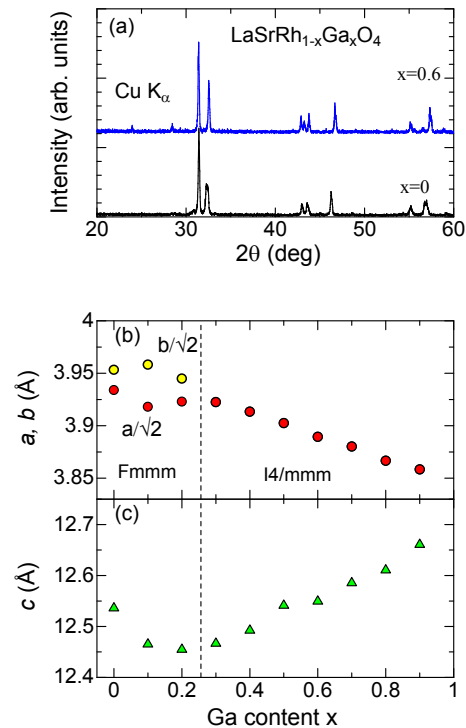}
 \caption{(color online)
(a) X-ray diffraction patterns of LaSrRh$_{1-x}$Ga$_x$O$_4$
 (x=0 and 0.6).
 (b)(c) Lattice parameters plotted as a function of the Ga content $x$.
 The space group changes from orthorhombic (Fmmm) to tetragonal 
 (I4/mmm)  between $x=$0.2 and 0.3.
 }
\label{fig1}
\end{figure}

\section{results and discussion}
Figure 1(a) shows typical x-ray diffraction patterns for the prepared
samples.
We find that LaSrRhO$_4$ ($x=0$) and LaSrGaO$_4$ ($x=1$) 
make a solid solution in the whole range of $x$.
All the peaks for $x=0$ are indexed as  the orthorhombic structure with the space
group Fmmm, as is consistent with preceding
papers.\cite{blasse1965,shimura1994}
With increasing Ga content $x$, the symmetry changes from orthorhombic to
tetragonal around $x=0.3$, as is evidenced by the small peaks around 
$2\theta$=23 and 28 deg for $x=0.6$.
This tetragonal structure is consistent with the other end phase of
LaSrGaO$_4$ with the space group of I4/mmm.\cite{britten1995}
Possibly owing to the symmetry change, the $c$ axis length 
shown in Fig. 1(c) takes a minimum around $x=0.3$, 
while the $a$ and $b$ axis lengths 
shown in Fig. 1(b) rather smoothly decrease with $x$.
Since  a Ga$^{3+}$ ion has a smaller ionic radius than a Rh$^{3+}$ ion,
the lattice volume smoothly decreases with $x$ (not shown).

\begin{figure}
 \includegraphics[width=7cm,clip]{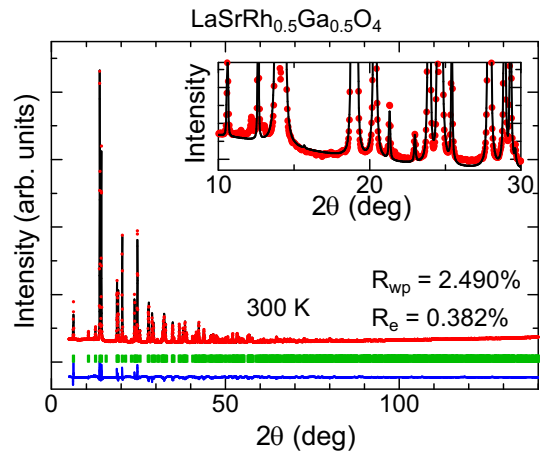}
 \caption{(color online)
Rietveld refinement for the 
x-ray diffraction patterns of LaSrRh$_{0.5}$Ga$_{0.5}$O$_4$.
The inset shows an extended figure from $2\theta$= 10 to 30 deg.
Although a tiny amount of unindexed peak is seen around 12 deg,
the refinment excellently reproduces the observed data. }
\label{fig2}
\end{figure}

The title compound belongs to the Ruddlesden-Popper series 
of (La,Sr)$_{n+1}$(Rh,Ga)$_{n}$O$_{3n+1}$, and 
$n\ne 1$ phases often grow as a secondary phase.
In the present case, (La,Sr)(Rh,Ga)O$_3$ ($n = \infty$)can grow in the same preparation
conditions, and a small amount of such impurity may
change the valence of the rhodium ion from $3+$.
In order to check this possibility, we have measured synchrotron x-ray
diffraction for the $x=0.5$ sample.
Figure 2 shows the synchrotron x-ray diffraction
pattern at room temperature.
There are some unindexed reflections around 12 deg, 
but their intensity is less than 0.6\% of the main peak.
We performed the Rietveld refinement, and find that 
the resultant fitting is reasonably well.
Thus we safely conclude that the prepared powder samples 
are stoichiometric and in single phase within an uncertainty of less
than 0.6\%.

Next we evaluate the valence of the rhodium ion 
from the transport properties.
Figure 3(a) shows the resistivity 
of LaSrRh$_{1-x}$Ga$_x$O$_4$
plotted as a function of inverse temperature.
The room-temperature resistivity is 1 $\Omega$cm for $x=0$,
which is higher than the previously reported 
data by Shimura et al.\cite{shimura1994}
Figure 3(b) shows the Seebeck coefficient 
of LaSrRh$_{1-x}$Ga$_x$O$_4$
plotted as a function of inverse temperature.
Again, the room-temperature thermopower for $x=0$ is
larger than the data  by Shimura et al.\cite{shimura1994}
These results indicate that the carrier concentration
of our sample is lower than that of their samples.
The positive sign of the Seebeck coefficient
indicates that the valence of the
rhodium ion is larger than 3.
Considering the fraction of the impurity phase is less than 0.6\%,
the valence of the rhodium ions ranges from 3.00 to 3.02.
Note that the Heikes formula \cite{chaikin1976} is not valid
in the present case because of significant temperature variation.
The isostructual LaSrCoO$_4$ shows nearly the same
value of 200 $\mu$V/K at 300 K with strong temperature variation.
\cite{chichev2006}

Let us have a closer look at the compositional dependence
of the transport data.
The resistivity increases with increasing Ga content, and 
the slope in the Arrhenius plot increases concomitantly.
The energy gap ($E_g^{\rho}$) determined by the slope is plotted
in the inset of Fig. 3(b);
the magnitude is 0.1--0.3 eV, which is typical for
insulating transition-metal oxides.
Since the $x=1$ sample has a band gap larger than 3 eV, 
the increase in $E_g^{\rho}$ indicates a systematic evolution
of the electronic states.
The Seebeck coefficient increases with decreasing temperature
around room temperature, suggesting an activation-type transport.
By evaluating the slope of the Seebeck coefficient against $1/T$
shown by the dotted lines,
we determine the energy gap ($E_g^S$) as is also plotted
in the inset of Fig. 3(b).
$E_g^{\rho} \gg E_g^S$ for $x< 0.5$
indicates that the activation energy predominantly comes from
the mobility, and the system is well described in terms of 
small polaron.\cite{palstra1997}
On the contrary, $E_g^{\rho} \sim E_g^S$ for $x=0.6$ 
indicates that the activation energy comes from the 
energy gap in the density of states.
This indicates that the system continuously evolves
from a small polaron insulator to a band insulator.
We further note that the unwanted holes are 
negligible for $x>0.5$ at low temperatures because
of the gap in the density of states.

\begin{figure}
 \includegraphics[width=8.5cm,clip]{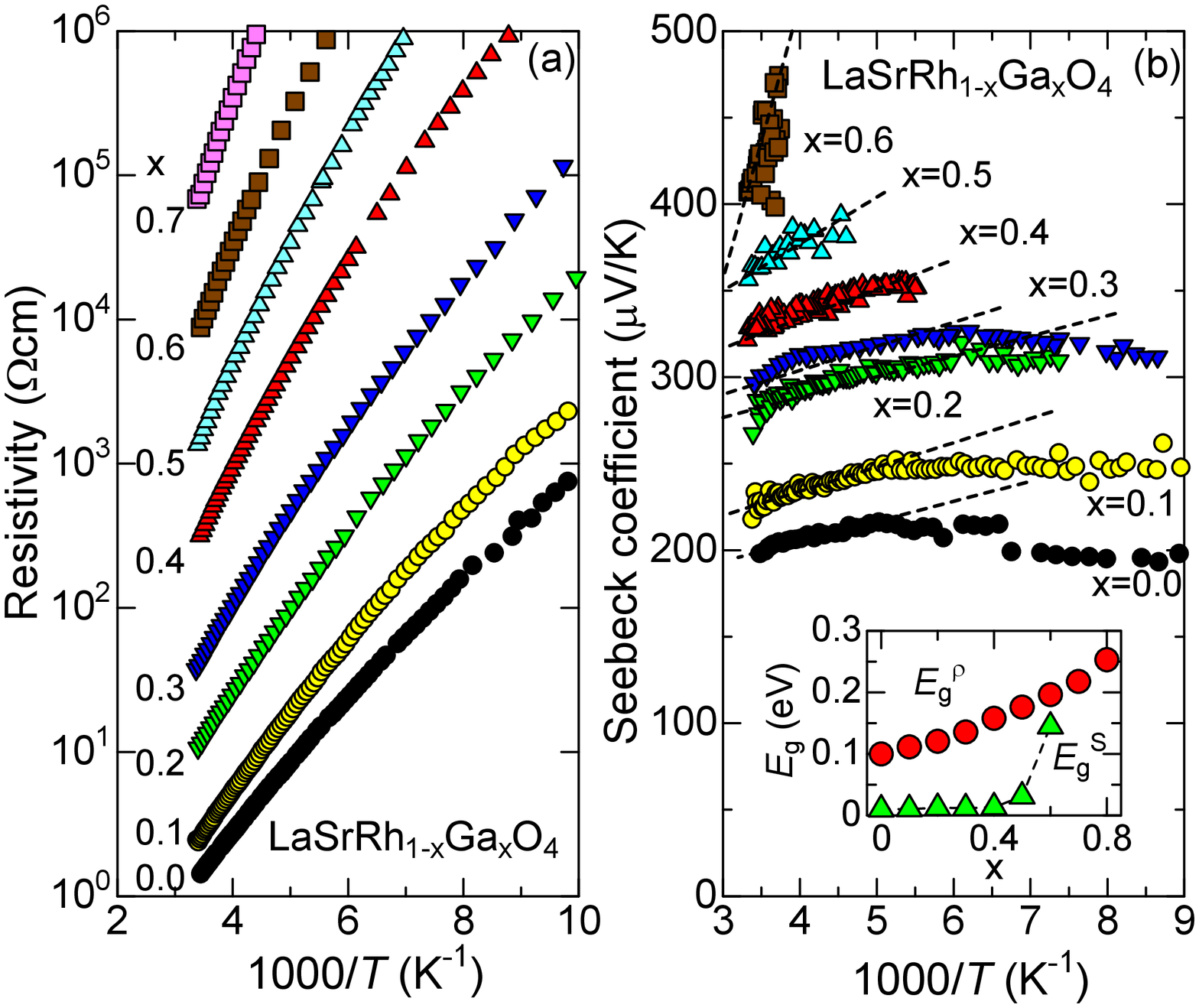}
 \caption{(color online)
(a) Resistivity and (b) the Seebeck coefficients of the prepared samples.
The dotted lines in (b) are guide to the eye.
The inset in (b) shows the energy gap evaluated from the resistivity
($E_g^\rho$) and the Seebeck coefficient ($E_g^S$).
 }
\label{fig3}
\end{figure}

Now we focus on the magnetism of our samples.
Figure 4 shows the susceptibility of LaSrRh$_{1-x}$Ga$_x$O$_4$.
A first thing to note is that \textit{all the samples are paramagnetic}.
Considering that Sr$^{2+}$, La$^{3+}$ and Ga$^{3+}$ ions are
diamagnetic, we have come to the conclusion that Rh$^{3+}$ is 
\textit{magnetic},
which is seriously incompatible with our collective wisdom.
A second feature is that all the data are roughly inversely proportional
to temperature, 
suggesting that the magnetic moment on the  Rh$^{3+}$ ion
is independent from each other.
A third feature is that the temperature-independent susceptibility 
is significant, and changes its sign with $x$. 
We also emphasize that the susceptibility of the cubic Rh$^{3+}$ oxide
LaRh$_{0.5}$Ga$_{0.5}$O$_3$ shows a much smaller 
paramagnetic signal.
This indicates that the paramagnetic response 
of LaSrRh$_{1-x}$Ga$_x$O$_4$  is inherent 
in the layered structure of $A_2B$O$_4$, 
and does not come from
impurities in the raw powdered oxides of Rh$_2$O$_3$,
La$_2$O$_3$, and Ga$_2$O$_3$.

\begin{figure}
\includegraphics[width=7cm,clip]{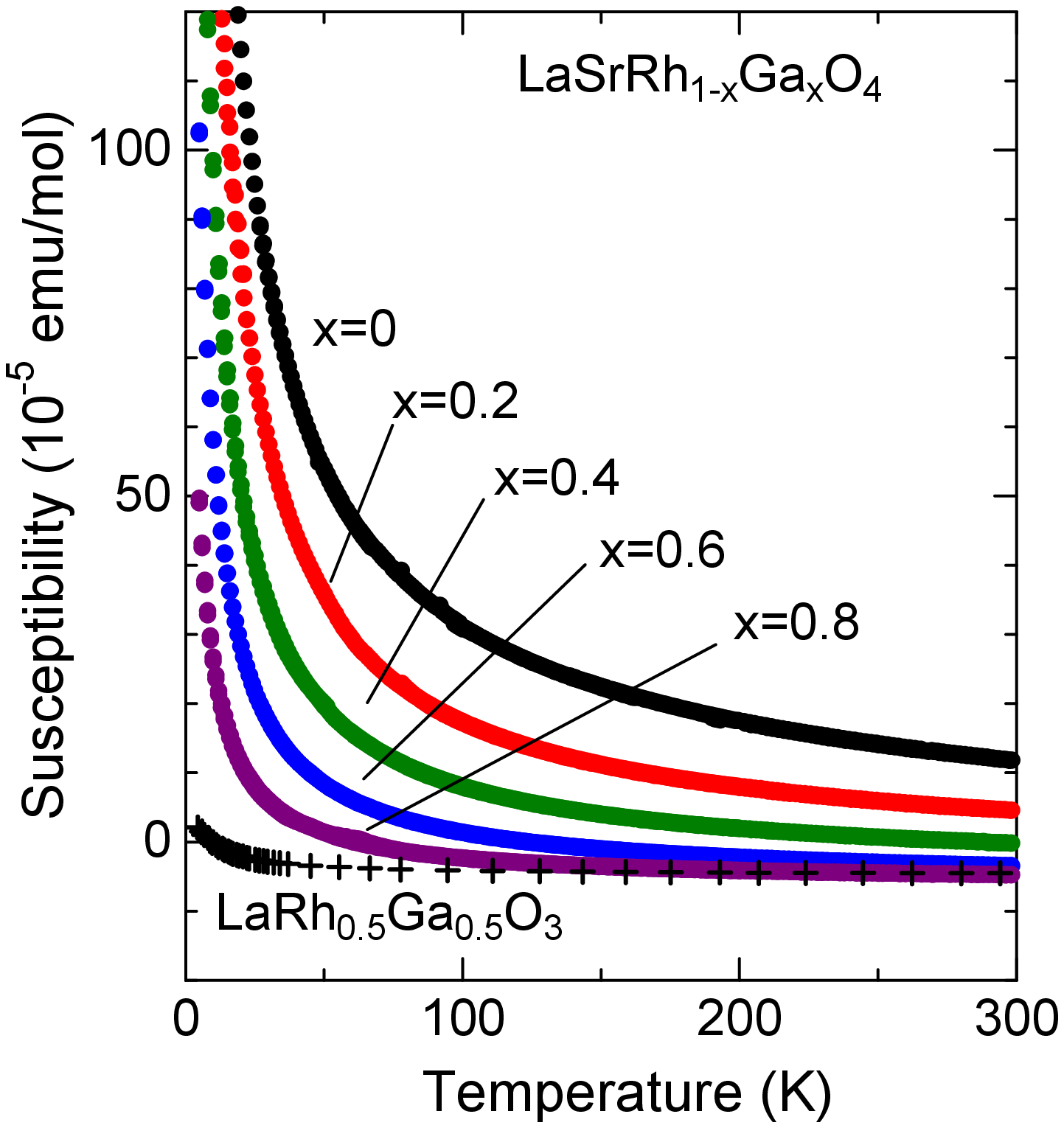} 
\caption{(color online) 
Magnetic susceptibility of  LaSrRh$_{1-x}$Ga$_x$O$_4$.
All the data exhibits Curie-Weiss-like paramagnetism.
The data for the cubic Rh oxide LaRh$_{0.5}$Ga$_{0.5}$O$_3$ is also
plotted. The paramagnetic signal is far smaller, indicating that
the paramagnetism is inherent in the layered structure.}

\vspace*{5mm}
\includegraphics[width=8cm,clip]{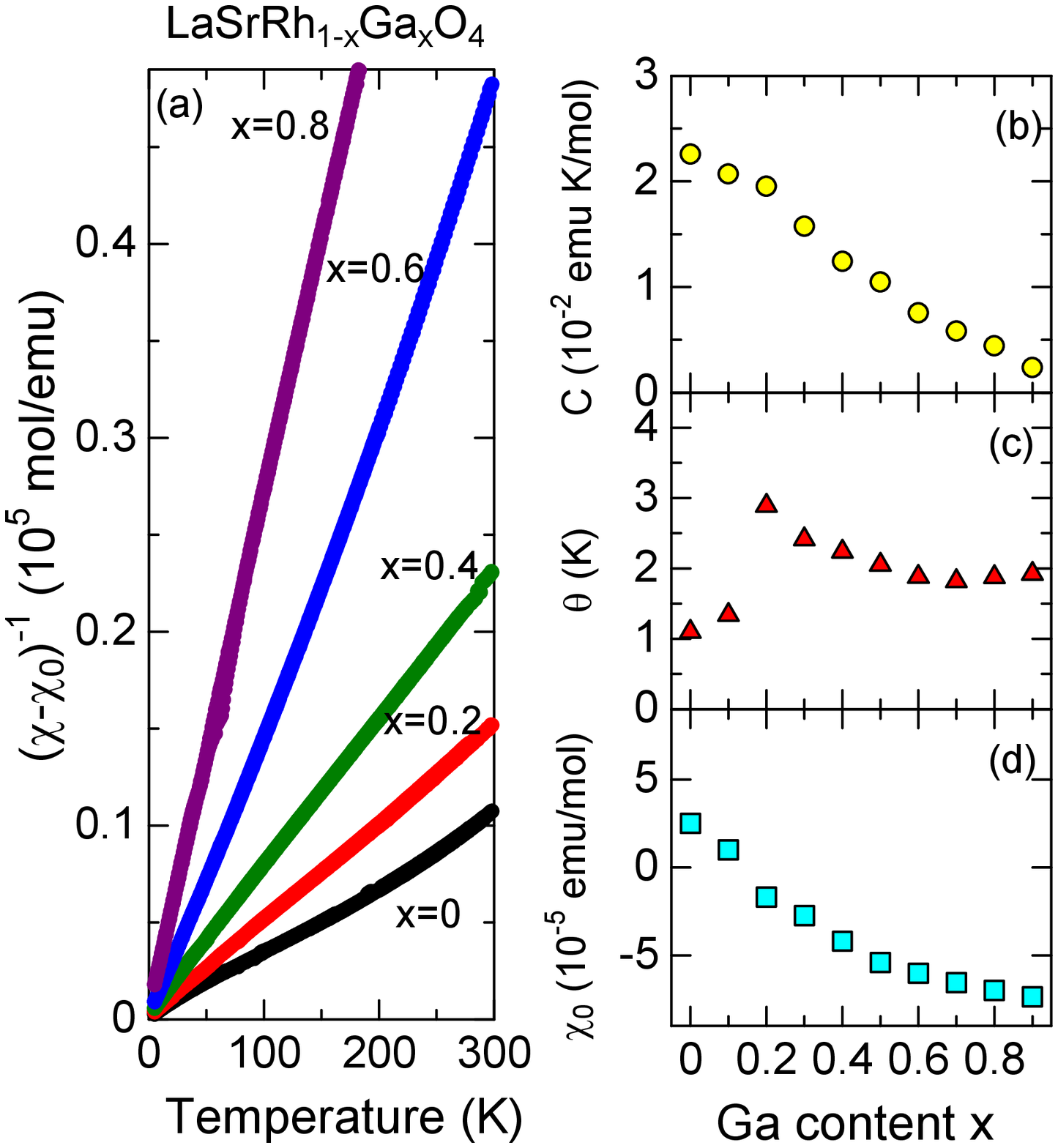} 
\caption{(color online) 
 (a) Inverse susceptibility of LaSrRh$_{1-x}$Ga$_x$O$_4$
 plotted as a function of temperature. 
Note that the temperature-independent susceptibility
$\chi_0$ is subtracted.
 (b)(c)(d) The parameters obtained from the susceptibility 
 in Fig. 4 by fitting with the Curie-Weiss law. (b) The Curie constant
(c) the Weiss temperature and (d) the temperature independent 
 susceptibility $\chi_0$.
}
\label{fig4}
\end{figure}

Considering the above features, we fit the experimental data with 
a modified Curie-Weiss law given by
  \begin{equation}
   \chi = \frac{C}{T+\theta}+\chi_0,
  \end{equation}
where $C$, $\theta$ and $\chi_0$ are the Curie constant,
the Weiss temperature, and the temperature-independent susceptibility,
respectively.
As shown in Fig. 5(a),
the inverse susceptibility $(\chi-\chi_0)^{-1}$ is found to 
be linear in $T$ down to the lowest temperature measured.
Figures 5(b)-5(d) show the fitting parameters.
As expected, the Weiss temperature is determined to be a small value
of the order of 1 K.
Thus, except for low temperatures, the spin-spin interaction can be
neglected.
The Curie constant is 0.02 emu K/mol for $x=0$, which is 
3\% of that observed in 
LaSrCoO$_4$.\cite{chichev2006, moritomo1997, shimada2006, ang2008}
This implies that only 3\% of the Rh ions are magnetic.
It should be emphasized that all the Curie constants decrease almost
linearly with $x$ [Fig. 5(b)], which implies that 
the Ga substitution simply causes a dilution effect,
and the magnetic Rh ions are always 3\% of the Rh ions 
for all the samples.
We should emphasize that the fraction of 3\% 
is much larger than the purity of the raw-material powders and
the volume fraction of impurity phases  evaluated above.
In addition, we can neglect thermally activated Rh$^{4+}$ ions 
for $x>0.5$ at low temperature because of the finite energy gap 
in the density of states (the inset of Fig. 3),
and yet observe the Curie-Weiss behavior.

\begin{figure}
\includegraphics[width=8cm,clip]{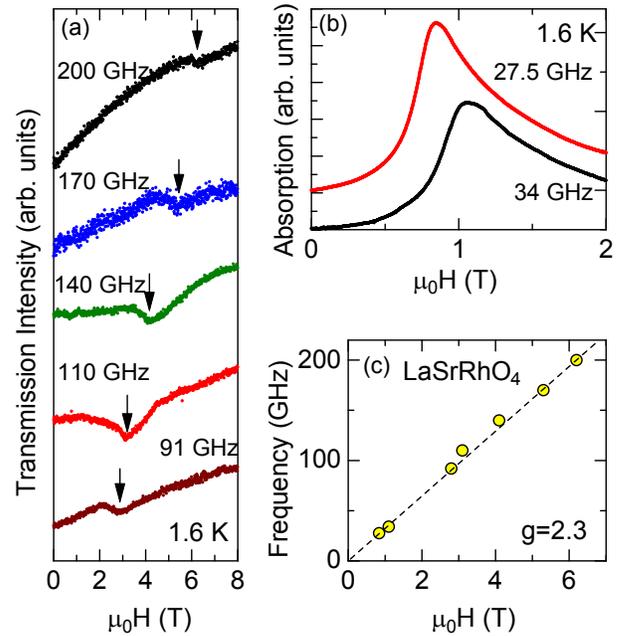}
\caption{(color online) 
 (a) Millimeter-wave adsorption by electron spin resonance 
 measured with a non-resonant transmission at 1.6 K
 in LaSrRhO$_4$.
 The arrows indicate the resonance field.
 (b) Microwave absorption using cavity perturbation with 
 resonance frequencies of 27.5 and 34 GHz.
 (c) Resonance frequency plotted as a function of external field.
 From the slope, the $g $ value is evaluated to be 2.3.
}
\label{fig5}
\end{figure}

Figure 6(a) shows the millimeter-wave transmission intensity
of LaSrRhO$_4$ plotted as a function of external field $\mu_0H$
at 1.6 K.
As indicated by the arrows,  all the transmission curves have
a broad dip, which corresponds to electron spin resonance 
at the field.
The field at which the dip is observed increases 
with increasing frequency. 
The dip width is as large as 1 T, 
suggesting a short spin-lattice relaxation time in this system.
Figure  6(b) shows the absorption curve measured
with resonant cavities for 27.5 and 34 GHz at 1.6 K.
An absorption peak is clearly visible near 1 T with a broad width of 1 T.
We should note that the shape of the absorption curve is
not symmetric to the resonance field.
Such an asymmetric shape has been analysed 
with a Dysonian function,\cite{dysonian}
but the fitting was not satisfactory for the present data (not shown).
At present, we do not understand the origin for 
the absorption asymmetry, but we speculate that 
the resonance condition seems to change with
increasing external fields, which implies that the dielectric
constant may depend on magnetic field.

Figure 6(c) shows the resonance frequency plotted as a function
of resonance field.
As is clearly seen, the frequency $\nu$ is linear in magnetic field $\mu_0H$
within experimental errors.
This is indeed what is expected in electron spin resonance for
a noninteracting spin system,
and is consistent with the very small $\theta$ in Fig. 5(c).
We obtain the $g$ value from the proportionality constant
expressed by $h\nu = g \mu_B\mu_0H$ to be $g=$2.3.
This value indicates that orbital angular moment $L$ is quenching.
The perovskite oxide LaCoO$_3$ shows a $g$ value of 3.35, \cite{nojiri2002}
which is explained in terms of $L=1$ in $t_{2g}$ orbitals.
Thus the value of $g$=2.3 excludes the possibility
that the Rh$^{3+}$ is in  the high-spin state.


\begin{figure}
 \includegraphics[width=6cm,clip]{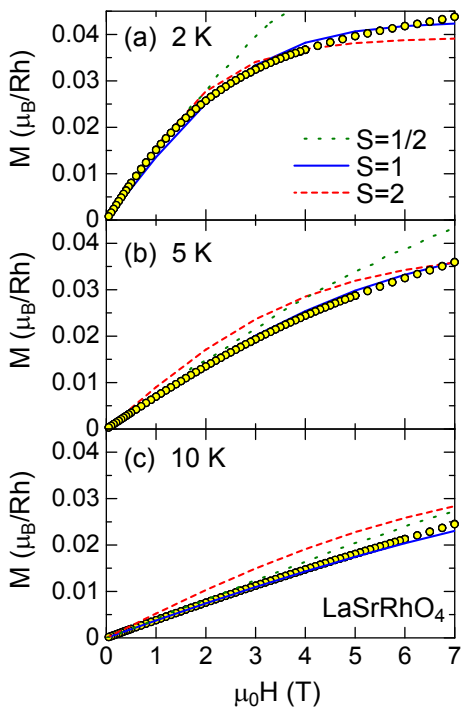}
\caption{(color online) 
 Magnetization-field curves of LaSrRhO$_4$
 fitted with a modified Brillouin function
 (see text). (a) 2 K, (b) 5 K and (c) 10 K,
 The open circles represent the measured magnetization, and the solid,
 broken and  dotted curves are numerically calculated curves 
 for $S=1$, $S=2$ and $S=1/2$, respectively.
The calculated curves are drawn to fit 
the  low-field magnetization at 2 K.
}
\label{fig6}
\end{figure}

Next let us examine the spin state of the Rh ions responsible
from the field dependence of the magnetization.
Since the electron configuration of Rh$^{3+}$
is $(4d)^6$, a possible magnetic state is the intermediate-spin state
($S$=1) or  the high-spin state ($S$=2).
Thanks to the weak spin-spin interaction,
we can employ the Brillouin function for the fitting,
with which we can distinguish $S=1$ from $S=2$.
Figure 7 shows the magnetization-field curves
of LaSrRhO$_4$ taken at 2, 5 and 10 K.
The magnetization shows saturation behaviour in
high fields, which is  more significant at lower temperature.
This  is qualitatively the same as is expected from
the Brillouin function.
Considering the small value of $\theta$, we slightly 
modify the Brillouin function to include the magnetization $M$
from the spin-spin interaction, and propose
a following function given by
\begin{equation}
 B_S^*(H, M) = \frac{\sum -g\mu_BS_z
  \exp\left[-\beta g\mu_BS_z\mu_0(H-\alpha M)\right]}
  {\sum\exp\left[-\beta g\mu_BS_z\mu_0(H-\alpha M)\right]},
\end{equation}
where $\alpha$ is the molecular field coefficient given by
$\alpha = \theta/C$, and $\beta$ is the inverse temperature
$\beta = 1/k_BT$.
Thus the magnetization $M$ is determined by the following
self consistent equation expressed as
\begin{equation}
 M = fN_0B_S^*(H,M) +\chi_0 H, 
\end{equation}
where $f$ is the fraction of the magnetic Rh$^{3+}$ ions.
We take $\alpha$ and $\chi_0$ from the data in Fig. 5,
and $g=2.3$ from Fig. 6.
Consequently, the fraction $f$ is left as the only one adjustable parameter.
The solid, broken and dotted curves in Fig. 7 represent
the calculations for $S=1$ , $S=2$, and $S=1/2$ respectively.
The adjustable parameter $f$  is taken to be 1.9 \% for $S=1$,
0.85\% for $S=2$, and 4.7\% for $S=1/2$,
in order to fit low-field magnetization at 2 K. 
As is clearly seen in Fig. 7, the $S=1$ curves consistently
explain the measured magnetization.
The large deviation of the calculated $S=1/2$ curve
clearly excludes a possibility that some Rh$^{3+}$ ions
may be disproportionated as Rh$~{2+}$ and Rh$^{4+}$ 
to work as $S=1/2$.

\begin{figure}
 \includegraphics[width=7cm,clip]{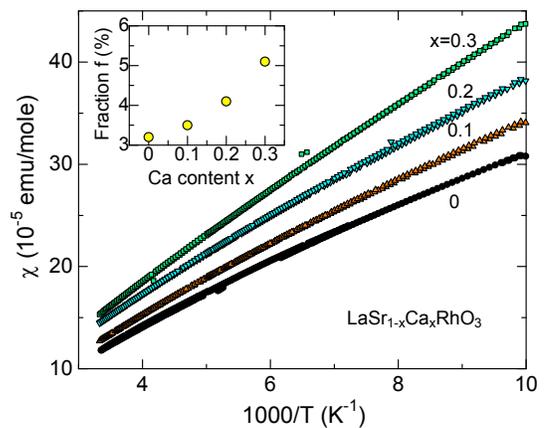}
\caption{(color online) 
 Invese  susceptibility $(\chi-\chi_0)^{-1}$
 of LaSr$_{1-x}$Ca$_x$RhO$_4$
 plotted as a function of temperature. 
 The temperature-independent susceptibility $\chi_0$
 is 1.8, 1.5, 1.3 and 0.8$\times$10$^{-5}$ emu/mol
for $x=$0, 0.1, 0.2 and 0.3, respectively.
 The slope changes significantly with the Ca content $x$.
 The intermediate spin state for the tetragonally distorted
 octahedra is schematically shown in the inset.
 Fraction of the magnetic Rh$^{3+}$ ions 
 evaluated from the slope
 is shown in the inset.
}
\label{fig7}
\end{figure}

In order to examine the $A$-site disorder effect,
we prepared a set of samples of LaSr$_{1-x}$Ca$_x$RhO$_4$.
Figure 8 shows the inverse susceptibility  $(\chi-\chi_0)^{-1}$
of LaSr$_{1-x}$Ca$_x$RhO$_4$  plotted as a function of
temperature.
 The temperature-independent susceptibility $\chi_0$
 is 1.8, 1.5, 1.3 and 0.8$\times$10$^{-5}$ emu/mol
for $x=$0, 0.1, 0.2 and 0.3, respectively.
Unlike the susceptibility of LaSrRh$_{1-x}$Ga$_x$O$_4$,
the slope of the susceptibility increases with the Ca content $x$.
By fitting the susceptibility with Eq. (1), we obtain the Curie constant
$C$, from which we evaluate the fraction of the magnetic Rh$^{3+}$
through the relation as
\begin{equation}
 C = \frac{fN_0g^2S(S+1){\mu_B}^2}{3k_B}.
\end{equation}
By putting $g$=2.3 and $S$=1, we get the fraction $f$ as shown in
the inset of Fig. 8.
We notice that for $x$=0,  the fraction of 3.1 \% evaluated
from Eq. (4) is slightly larger than 1.9 \% evaluated from Eq. (3).
Normally, the $g$ value is determined by ESR accurately, 
but the distorted signal makes it impossible this time. 
Since the $g$ value is determined to be 2.3$\pm$0.1, 
the disagreement in the fraction  may come 
from the ambiguity in the $C$ value.
An important feature is that the fraction increases with 
the Ca content $x$, which implies that the magnetic Rh$^{3+}$ ions
are related to the degree of the $A$-site disorder.

Finally let us discuss a possible origin of the magnetic Rh$^{3+}$ ions
distributed with a fraction of 2-5\%.
A first point is that their magnetic moment is
stable at all temperature measured, and the number of the magnetic 
ions is independent of temperature, for the susceptibility
obeys the Curie-Weiss law with a small Weiss temperature
in a wide range of temperature. 
A second point is that the magnetic Rh$^{3+}$ is local, just like a magnetic
impurity, and  the fraction is almost independent of the $B$-site disorder,
but depends on the $A$-site disorder.
A third point is that such magnetic signal is related to
the layered structure, possibly related to the tetragonal distortion
of the RhO$_6$ octahedron.
Note that the susceptibility of LaRh$_{0.5}$Ga$_{0.5}$O$_3$,
has a negligibly small Curie tail as shown in Fig. 4.

\begin{figure}
 \includegraphics[width=6.5cm,clip]{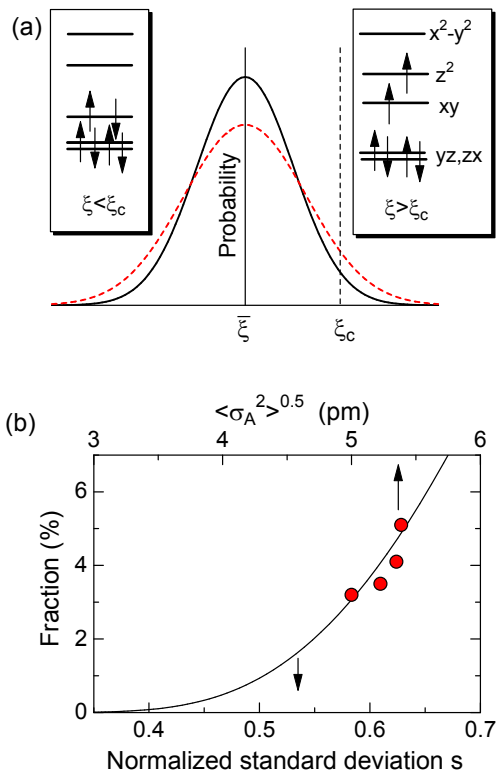}
\caption{(color online) 
(a) A possible distribution of the bond length $\xi$  of 
 the Rh ion and the apical oxygen ion.
 When $\xi$ exceeds a critical value of $\xi_c$,
 the energy gap of $xy$ and $z^2$ becomes
small enough to stabilize the intermediate spin state.
For $\xi <\xi_c$, the low spin state is favored.
The fraction of the intermediate spin is
larger in a more disordered distribution (the dotted curve).  
(b) The theoretical curve for the fraction plotted
as a function of the normalized standard deviation $s$
given by Eq. (11).
The observed fraction is also plotted as a function of 
 the $A$-site disorder $\langle\sigma_A^2\rangle  
\equiv \langle {r_A}^2 \rangle - \langle r_A \rangle^2$,
 where $r_A$ is the ionic radius of the $A$ site ions (see text).
}
\label{fig8}
\end{figure}

Attfield et al. have discovered that the $A$-site
disorder seriously affects the superconducting transition temperature
in the doped La$_2$CuO$_4$.\cite{attfield1998}
Since the transition temperature is sensitive to the bond length 
of Cu$^{2+}$ and apical O$^{2-}$ ions,\cite{ohta1991}
the variance in the out-of-plane Cu-O distance 
may deteriorate the superconducting properties.
We apply a similar story to the title compound.
When the Rh-O bond length $\xi$ 
along the $c$ axis direction is sufficiently short, 
we can regard RhO$_6$ as a regular
octahedron, where the low spin state is stable 
(the left schematic in Fig. 9(a)).
On the other hand, when $\xi$ exceeds a critical value of $\xi_c$,
a long $\xi$ largely splits the energy level between $x^2-y^2$
and $z^2$, and eventually shifts  the $z^2$ level downwards. 
Likewise, it also largely splits the energy level between $xy$
and $yz/zx$, and raises the $xy$ level.
As a result, a small energy gap between the $z^2$ and $xy$ 
levels will favor the spin state shown in the right schematic in Fig. 9(a).
This configuration is identical to the intermediate spin state proposed 
for LaSrCoO$_4$.\cite{wang2005} 

We will roughly estimate how the fraction $f$
of the magnetic moment is related to the disorderness of the A-site cation.
Assuming a Gaussian distribution with the mean value  
$\bar \xi$ and the variance $\sigma^2$,
$f \equiv f(\bar\xi, \sigma^2)$ equals to the probability for $\xi > \xi_c$ given by
\begin{equation}
 f(\bar\xi, \sigma^2)=
 \frac{1}{\sqrt{2\pi\sigma^2}}\int_{\xi_c}^{\infty}
 \exp\left( -\frac{(x-\bar\xi)^2}{2\sigma^2} \right) dx.
\end{equation}
By replacing the variable $t=(x-\bar\xi)/\sqrt{2\sigma^2}$, we get
\begin{eqnarray}
 f(\bar\xi, \sigma^2) &=& \frac{1}{\sqrt{\pi}}
  \int_{\Delta\xi/\sqrt{2\sigma^2}}^{\infty} \exp\left(-t^2 \right) dt\\
 &=& 2 {\rm erfc}\left(\frac{\Delta\xi}{\sqrt{2\sigma^2}} \right),
\end{eqnarray}
where  erfc$(t)$ is the complementary error function, 
and $\Delta\xi=\xi_c-\bar\xi$.
If we assume $\bar\xi$ to be independent of the Ca concentration
in LaSr$_{1-x}$Ca$_x$CoO$_4$,
we find $s\equiv \sqrt{2\sigma^2}/\Delta\xi$ is the only parameter,
and the fraction is described simply as
\begin{equation}
 f(s) =  2 {\rm erfc}\left(\frac{1}{s}\right).
\end{equation}
Figure 9(b) depicts such relationship,
where the fraction $f(s)$ is plotted from $s=$0.35 to 0.7.
One can see that $f(s)$ is a monotonically increasing function of $s$,
and the faction increases with the variance in the Rh-O bond.
We also plot the standard deviation in the A-site ionic radius as
$\sqrt{\sigma_A^2} = \sqrt{\langle {r_A}^2 \rangle - \langle r_A \rangle^2}$
in the same graph.
Although this quantity does not
equal the standard deviation in the Rh-O distance, one can see 
that $\sqrt{\sigma_A^2}$ roughly follows the curve given by Eq. (8).

\section{summary}
 We have prepared a set of polycrystalline samples of 
LaSrRh$_{1-x}$Ga$_x$O$_4$ and LaSr$_{1-x}$Ca$_x$RhO$_4$,
 and have measured the resistivity, Seebeck coefficient,
magnetization and electron spin resonance
 in order to evaluate the magnetic properties and spin states
 of the layered rhodium oxides.
 We find that all the samples show Curie-Weiss-like susceptibility with 
a small Weiss temperature of the order of 1 K,
which is seriously incompatible with the collective wisdom
that a trivalent rhodium ion is nonmagnetic.
The $g$ factor is determined to be  $g=$2.3 from 
the electron spin resonance, and the spin number is determined as $S=1$
from the magnetization-field curves by fitting with a modified 
Brillouin function.
The fraction of the $S=1$ spins is 2-5\%, and 
the disorder in the La/Sr/Ca-site determines the spin fraction.

This work was partially supported by the collaboration with NGK
Insulators Ltd., and by ALCA, Japan Science and Technology Agency.
The synchrotron x-ray diffraction was performed under the approval of
the Photon Factory Program Advisory Committee (Proposal No. 2009S2-S008).


%

\end{document}